\documentclass[twocolumn,showpacs,preprintnumbers,amsmath,amssymb]{revtex4}

\usepackage{graphicx}
\usepackage{dcolumn}
\usepackage{bm}

\begin{document}


\title{Effective field theory with a 
$\theta$-vacua structure for 2d spin systems
}

\author{Akihiro Tanaka}
\author{Xiao Hu}
\affiliation{%
Computational Materials Science Center, National Institute for Materials Science, Sengen 1-2-1, Tsukuba, Ibaraki, 305-0047, Japan
}%

\date{\today}

\begin{abstract}
We construct a nonlinear sigma (NL$\sigma$) model description of 2+1d spin systems, 
by coupling together antiferromagnetic 
spin chains via interchain exchange terms. 
Our mapping incorporates methods developed recently by ourselves and by Senthil and Fisher, 
which aim at describing competition between antiferromagnetic and valence-bond-solid orders  
in quantum magnets. 
The resulting 2+1d $O$(4) NL$\sigma$ model 
contains a topological $\theta$-term whose vacuum angle $\theta$ varies continuously with $\delta$, 
the bond-alternation strength of the interchain interaction. This implies that the 
$\theta$-vacua structure for this NL$\sigma$ model can be explored by tuning $\delta$ in a suitable 
2d spin system, which is strongly reminiscent of the situation for 1+1 AF spin chains with bond-alternation.
\end{abstract}

\maketitle


Haldane's semiclassical description of antiferromagetic (AF) spin chains 
in terms of an $O$(3) nonlinear sigma (NL$\sigma$) model   
is a simple and yet powerful format for     
detecting  
quantum exotica in 1d\cite{Fradkin's-book}. 
In this language one exercises special care to 
keep track of possible ^^ ^^ $\theta$-term" \cite{high-energy} type Berry phases, 
which introduces complex-number valued weights, and hence nontrivial 
quantum interference into the the path-integral representation of the partition function, 
$Z[{\boldsymbol n}]=\int{\cal D}{\boldsymbol n}e^{-(S_{NL\sigma}+S_{\theta})}$.   
Here the $\theta$-term action is 
$S_{\theta}=i\theta Q_{\tau x}$, where the winding number 
${\cal Q}_{\tau x}\equiv\frac{1}{4\pi}\int d\tau dx 
{\boldsymbol n}\cdot \partial_{\tau}{\boldsymbol n}\times\partial_{x}{\boldsymbol n}
\label{Q_{taux}}$
probes the global topology of the unit N{\'e}el vector ${\boldsymbol n}$'s space-time configuration.  
Hereafter we refer to the coefficient $\theta$ as the vacuum angle.   
The presence of $S_{\theta}$ has profound consequences,  
altering nonperturbatively the system's ground state properties. 
In particular, the system undergoes a continuous quantum phase transition 
when the vacuum angle traverses the value $\theta=\pi$ modulo $2\pi$.  
Following the success in discriminating between the  
spectral properties of integer and half-integer spin systems
\cite{Fradkin's-book}, 
this aproach was applied to bond-alternating AF Heisenberg spin chains\cite{Early-works}, 
described by the Hamiltonian 
$H=\sum_{i}J(1-(-1)^i \delta){\boldsymbol S}_{i}\cdot{\boldsymbol S}_{i+1}$. 
The vacuum angle was found to 
depend on the bond-alternation strength $\delta$ as 
$\theta=2\pi S(1-\delta)$. 
Sweeping $\delta\in[-1,1]$ 
would therefore enable one to probe the entire $\theta$-vacua structure.  
Physically, this implies the existence of $2S$ successive 
quantum phase transitions between different valence-bond-solid (VBS) states, 
as later verified by numerical studies\cite{Numerics}.                                                                                                                                                  
This mapping technique is now incorporated routinely to 
determine the global phase diagram of various 1d and quasi-1d AF systems. 

The present study is an attempt to construct a NL$\sigma$ model 
exhibiting a nontrivial $\theta$-vacua structure, which describes 
a class of {\it bulk 2d spin systems}. 
This is in part motivated by
recent interest in exotic phase transitions in 2d AFs, e.g. between 2d VBS states\cite{Ashwin}. 
Here we will show that certain inter-VBS transitions in 2d can, like in 1d, 
indeed be described by a NL$\sigma$ model 
with a $\theta$-term, albeit with the target manifold $O$(4). (Observe that four is the 
required number of components for constructing the winding number ${\cal Q}_{\tau xy}$  
(defined later), which is the 2+1d analogue of ${\cal Q}_{\tau x}$.)  
Searches for novel Berry phase terms within in the $O$(3) NL$\sigma$ model description of 
2d AFs were carried out in the past\cite{Fradkin's-book,Haldane-PRL-88,No-Hopf-term},  
which showed that singular hedgehog events  
can induce Berry phase factors that drive the system towards a VBS groundstate
\cite{Haldane-PRL-88,Read-Sachdev}.  
An alternative approach is to depart from the $O$(3) description 
where the amplitude of the AF order parameter is kept fixed, 
and to employ instead a composite 5-component order parameter explicitly 
representing the competition between 
the three AF and two (horizontal and vertical) VBS components. 
We found in a previous work starting from the $\pi$-flux mean-field anstaz\cite{Tanaka-Hu-PRL-2005}, 
that the resulting 2+1d $O$(5) NL$\sigma$ model 
contains a Wess-Zumino (WZ) term which reproduces the 
hedgehog Berry phase factors in certain limits. 
Here we follow this latter route.

We had observed in ref.\cite{Tanaka-Hu-PRL-2005} 
that ^^ ^^ freezing" the fluctuating VBS amplitude along one of the spatial 
directions, thereby reducing the number of effective components to four, 
will modify our effective $O(5)$ action into 
a 2+1d $O$(4) NL$\sigma$ model with a $\theta$-term.    
This would then suggest that a novel $\theta$-vacua structure arises in a system 
with a spatially anisotropic AF-VBS competition, a situation which might be realized e.g. 
in a coupled stack of 1d spin chains with sufficient intrachain AF-VBS competition.  
The direct relation to such systems is 
not totally apparent though, since we began with a spatially isotropic 2d system. 
For this reason we shall employ and extend a method recently suggested 
by Senthil and Fisher \cite{Senthil-Fisher}, 
also along similar lines, which starts from purely 1d systems. 
Upon coupling these system {\it uniformly} along a second direction, 
these authors arrive at the $O$(4) model with $\theta=\pi$. Below we will find that introducing a 
tunable {\it bond-alternation} 
along the transverse direction allows us to access {\it arbitrary} values of $\theta$, i.e. a 
$\theta$-vacua very similar to the spin chain problem mentioned above.

Our starting point is the level-1 1+1d $SU(2)$ Wess-Zumino-Witten (WZW) model 
(higher-level cases, i.e. higher-$S$ generalizations will be mentioned later). 
Here we collect features of this model which will shortly become necessary.       
Physically, the WZW action    
simply describes the quantum criticality arising from competing AF and VBS orders
~\cite{Tanaka-Hu-PRL-2002,Tanaka-Hu-PRL-2005,Senthil-Fisher}. 
This becomes transparent when one represents      
the $SU(2)$-valued field $g$ using a unit 4-vector $\phi_{\alpha}$ $(\alpha=0, 1, 2, 3)$, 
viz 
$g=\phi_{0}+i{\boldsymbol \phi}\cdot{\boldsymbol \sigma}$, where 
${\boldsymbol \phi}\equiv (\phi_1 , \phi_2 , \phi_3)$.  
Indeed, it is straightforward to show with the aid of nonabelian bosonization techniques,   
that $\phi_{0}$ and ${\boldsymbol \phi}$ 
each corresponds in spin language 
to the dimer (VBS) order parameter $(-1)^{i}<{\boldsymbol S}_{i}\cdot {\boldsymbol S}_{i+1}>$, and the N\'eel vector ${\boldsymbol n}(x)$
~\cite{Tsvelik's-book}.  
Expressed as a functional of the composite AF-VBS order parameter $\phi_{\alpha}$, the WZW action 
is an $O$(4) NL$\sigma$ model with a WZ term,  
\begin{equation}{\cal S}[\phi_{\alpha}]=\frac{1}{8\pi}\int d\tau dx\left[
(\partial_{\tau}\phi_{\alpha})^2 +(\partial_x \phi_{\alpha})^2 \right] +\Gamma[\phi_{\alpha}],
\end{equation}
where the WZ functional is 
\begin{equation}
\Gamma[\phi_{\alpha}]=\frac{i\epsilon^{abcd}}{\pi}\int_{0}^{1}dt\int d\tau dx
\phi_{a}\partial_{t}\phi_{b}\partial_{\tau}\phi_{c}\partial_{x}\phi_{d}.
\label{wzw term1}
\end{equation}
The extra parameter $t\in[0,1]$ is a 
complication common to all WZ-type terms 
(the simplest example being the Berry phase for a single spin); 
here it is used to locally frame the three-dimensional ^^ ^^ area"  
on the surface of the hypersphere $S_3$ swept out by the 
map $(x, \tau)\in S_{2}\rightarrow \phi_{\alpha}\in S_3$. 
We record two properties of $\Gamma[\phi_{\alpha}]$ 
which we incorporate below. First, the change under a slight deformation of the 
configuration $\phi_{\alpha}\rightarrow\phi_{\alpha}+\delta\phi_{\alpha}$ 
can be written as a {\it local} functional (i.e. one without $t$):  
\begin{eqnarray}
\delta \Gamma [\phi_{\alpha}(\tau, x)]
&=& \frac{i\epsilon^{abcd}}{\pi}\int d\tau dx
\phi_{a}\delta\phi_{b}\partial_{\tau}\phi_{c}\partial_{x}\phi_{d}
\nonumber\\
&&+O(\delta\phi_{\alpha}^2 ).
\label{variation of WZ term}
\end{eqnarray}
Secondly, since $\Gamma[\phi_{\alpha}]$ is linear in all four components of the vector $\phi_{\alpha}$, 
simultaneously flipping any three will induce a sign change. In particular, 
\begin{equation}
\Gamma[\phi_{0}, -{\boldsymbol \phi}]=-\Gamma[\phi_{0}, {\boldsymbol \phi}].
\label{sign of WZ term}
\end{equation}

Let us now follow ref.\cite{Senthil-Fisher} and stack our 1d 
$\phi_{\alpha}(\tau, x)$ chains along a second ($y$) spatial direction. 
We include a bond-alternation $\delta$ in the interchain coupling, anticipating 
its contribution to the $\theta$-term. 
As detailed below, the 
interchain interaction ${\cal S}_{\perp}=\int d\tau H_{\perp}$ 
required for our purpose of generating a family of $\theta$-vacua 
has the form, 
\begin{eqnarray}
H_{\perp}&=&-\int 
{\mbox {\hspace*{-1mm}}}
dx\sum_{y}[ J_{\perp}(1+(-1)^y {\delta}){\boldsymbol N}(x,y)\cdot{\boldsymbol N}(x,y+1)
\nonumber\\
&&
{\mbox {\hspace*{-5mm}}}
+J_{\perp}(1-(-1)^y {\delta})N_{0}(x,y)N_{0}(x,y+1)], 
\label{J_perp term}
\end{eqnarray}
with $J_{\perp}>0$, and  
$N_{\alpha}(x,y)=(N_{0}(x,y), {\boldsymbol N}(x,y))\equiv 
(\phi_{0}(x,y)+a(-1)^y l_{0}(x,y), (-1)^y {\boldsymbol \phi}(x,y)+a {\boldsymbol l}(x,y))$ ($a$ is the lattice constant). 
%
%
Here we have reinstated 
a small and rapidly fluctuating component $l_{\alpha}(x, y)=(l_{0}(x,y), {\boldsymbol l}(x,y))$,  
to be eventually integrated out, on top of the slowly 
varying field $\phi_{\alpha}(x, y)$. This intermediate step 
is easily understood when one recalls that  
for the 1d AF, the uniform magnetization ${\boldsymbol L}$ (the subdominant 
fluctuation in that case), playing an analogous role, needed to be integrated out to arrive at the 
final effective action\cite{Fradkin's-book}. 
To preserve the $SU(2)$ symmetry of the 
WZW field $g$, we need to impose the constraint $N_{\alpha}N_{\alpha}=N_{0}^2 +{\boldsymbol N}^2 \equiv 1$, which 
implies that $\phi_{\alpha}l_{\alpha}=0$. 
The form of $H_{\perp}$ dictates how 
the slow and rapid modes of the WZW fields on 
adjacent chains are to align; 
the slow fluctuations consist of a {\it staggered} alignment of ${\boldsymbol \phi}(x,y)$ fields and  
a {\it columnar} alignment of $\phi_{0}(x,y)$ fields along the $y$-direction~\cite{Senthil-Fisher}. 
This is consistent with our previous 
work, which suggested that an intrinsic competition between 
2d AF and columnar dimer states exists in the vicinity of the 
$\pi$-flux state~\cite{Tanaka-Hu-PRL-2005}. 
As for the rapidly varying modes, the vector ${\boldsymbol l(x,y)}$ 
tends to pile up in a ferromagnetic fashion,  
while the component $l_{0}$ corresponds to a staggered VBS configuration. 
This situation, depicted in fig.1, would be natural in the presence of a 
frustrating diagonal exchange interaction which is not dominantly strong.  
Later we will return to the issue of relating our study to a  
real spin system.  



\begin{figure}[htb]
\begin{center}
\end{center}
\caption{(Figure appended as a GIF file) 
Slowly and rapidly fluctuating degrees of freedom. Crossed lines indicate the presence 
of a frustrating diagonal exchange discussed in the text.}
\label{modes}
\end{figure}


Our goal below is to extract an effective theory for $\phi_{\alpha}(\tau,x,y)$.  
Readers familiar with Haldane's mapping for  
AF spin chains~\cite{Haldane-PRL-88} may find what follows more tractable by 
keeping in mind the set of correspondences between the 1d and 2d cases, displayed in Table 1. 
\begin{table}[h]
\caption{Correspondence with Haldane's mapping in 1d}
\begin{tabular}{cc} 
\hline
1d ( $O$(3) ) & 2d ( $O$(4) ) \\
\hline
${{\boldsymbol n}(\tau,x)+(-1)^x \frac{a}{S} {\boldsymbol L}(\tau,x)}$ & 
$\phi_{\alpha}(\tau,x,y)+a(-1)^y l_{\alpha}(\tau,x,y)$ \\
$iS\omega[{\boldsymbol n}(\tau,x)]$ & $\Gamma[\phi_{\alpha}(\tau,x,y)]$ \\
$i\delta \omega[{\boldsymbol n}]=i\int d\tau {\boldsymbol n}\cdot\delta{\boldsymbol n}\times\partial_{\tau}{\boldsymbol n}$ 
& eq.(\ref{variation of WZ term}) \\
$\omega[-{\boldsymbol n}]=-\omega[{\boldsymbol n}] $ & eq.(\ref{sign of WZ term}) \\
$S_{\theta}=i\theta{\cal Q}_{\tau x}$
& $S_{\theta}=i\theta{\cal Q}_{\tau xy}$
 \\            
\hline
\end{tabular}
\end{table}
In the second 
entry in the last row, 
${\cal Q}_{\tau xy}\equiv\frac{1}{2\pi^2}\int d\tau dxdy \epsilon^{abcd}
\phi_a \partial_{\tau}
\phi_b \partial_{x}\phi_c \partial_{y}\phi_d $, 
which is the winding number associated with the homotopy $\Pi_3 (S_{3})$,  
enters into the $\theta$-term for the 2+1d $O$(4) NL$\sigma$ model.

We now 
proceed to the continuum approximation. 
We begin by carrying out the $y$-summation over the WZ functionals 
with the help of 
eqs.(\ref{variation of WZ term}) and (\ref{sign of WZ term}),
\begin{eqnarray}
&&{\mbox {\hspace*{-8mm}}}
\sum_{y}\Gamma\left[
\phi_{0}+a(-1)^y l_{0}, (-1)^y {\boldsymbol \phi}+a{\boldsymbol l}\right]\nonumber\\
&&{\mbox{\hspace*{-6mm}}}
=
{\mbox {\hspace*{-1mm}}}
\sum_{y}(-1)^y\Gamma[\phi_{\alpha}]
+
\frac{ia}{\pi}\epsilon^{abcd}\sum_{y}
{\mbox {\hspace*{-2mm}}}
\int 
{\mbox {\hspace*{-1mm}}}
d\tau dx\phi_{a}l_{b}\partial_{\tau}\phi_{c}\partial_{x}\phi_{d}.
\label{sum over WZ functionals}
\end{eqnarray}
The alternating series in the last line 
is converted  
into an integral\cite{Senthil-Fisher},
\begin{equation}
\sum_{y}(-1)^y \Gamma[\phi_{\alpha}(y)]
\sim
\frac{1}{2}\int dy\partial_{y}\Gamma[\phi_{\alpha}]=-i\pi {\cal Q}_{\tau xy}.
\label{sum to integral}
\end{equation}
Next we turn to the interchain term $H_{\perp}$ which is decomposed into two contributions, 
$H_{\perp 1}$ and $H_{\perp 2}$, where
\begin{equation} 
H_{\perp 1}=-\int dxJ_{\perp}\sum_{y}N_{\alpha}(y)N_{\alpha}(y+1) ,
\end{equation} 
and 
\begin{eqnarray}
H_{\perp 2}&=&-\int dxJ_{\perp}\delta\sum_{y}(-1)^y [ N_{0}(y)N_{0}(y+1)
\nonumber \\
&&-{\boldsymbol N}(y)\cdot{\boldsymbol N}(y+1) ]. 
\end{eqnarray} 
Here we have suppressed the explicit dependence on $x$ for brevity. 
The continuum form of these interactions (discarding oscillatory contributions) read
\begin{equation}
H_{\perp 1}\sim\int dxdy \left(\frac{J_{\perp}}{2}a(\partial_{y}\phi_{\alpha})^2 
+2J_{\perp}al_{\alpha}^2 \right),
\label{H_perp1}
\end{equation} 
and
\begin{equation}
H_{\perp 2}\sim2J_{\perp}\delta a\int dxdy l_{\alpha}\partial_{y}\phi_{\alpha}.
\label{H_perp2}
\end{equation} 
Finally we integrate over ${l_{\alpha}}$,  
by collecting terms 
from eqs.(\ref{sum over WZ functionals}), (\ref{H_perp1}) and 
(\ref{H_perp2}), and completing the square with respect to $l_{\alpha}$. 
Dropping a higher-order derivative term and carrying out a suitable rescaling, 
our effective action reads 
\begin{eqnarray}
{\mbox{\hspace*{-3mm}}}
{\cal S}_{eff}[\phi_{\alpha}(\tau, x, y)]&=&-i\pi (1-\delta){\cal Q}_{\tau xy}
\nonumber \\
&&
{\mbox{\hspace*{-35mm}}}
+\int 
{\mbox{\hspace*{-1mm}}}
d^3 x
\frac{1}{2g}[\frac{1}{v}
(\partial_{\tau}\phi_{\alpha})^2 
+
{\mbox{\hspace*{-1mm}}}
 v
(\partial_{x}\phi_{\alpha})^2 
+
{\mbox{\hspace*{-1mm}}}
v
(1-\delta^2)(\partial_{y}\phi_{\alpha})^2 
],
\end{eqnarray}
where the velocity $v$ and the coupling constant 
$g$ each depends on parameters of the original Hamiltonian. 
The main findings here are (1) the dependence of 
the vacuum angle on the bond-alternation parameter, $\theta=\pi(1-\delta)$, and (2) the factor 
$(1-\delta^2 )$ which enters the coefficient for the interchain kinetic energy, both of which 
coincide with known results\cite{Early-works} for the S=1/2 bond alternated spin chain. 
In particular the 
vacuum angle $\theta=\pi$ (corresponding to $\delta=0$) lies at the point 
in parameter space where the 'strong' and 'weak' vertical 
bonds are interchanged. 
Further analogy with the $\theta$-vacua for the 1+1d case arises, 
when one generalizes the foregoing mapping to general $S$, by starting 
with the level $2S$ WZW model.  This gives $\theta=2\pi S(1-\delta)$, 
an exact reproduction of the 1d result mentioned earlier.
These strong similaritites naturally lead us to adopt the physical picture established 
within the spin chain context, and {\it associate 
the $\theta=\pi$  {\rm (}mod $2\pi${\rm )} points with phase transitions between different 
{\rm (}vertical{\rm )} VBS ground states}. 
(In contrast to the 1d problem, however, the precise nature (including the order of transition)  
of the 2+1d $O$(4) NL$\sigma$ model at $\theta=\pi$ 
so far does not seem to have been pursued.) 
 In the presence of an AF-favoring anisotropy, one can seek support for this expectation  
by viewing the problem in terms of the meron gas expansion\cite{high-energy,Meron} 
described (for $\theta=\pi$) in    
ref.\cite{Senthil-Fisher}. Extending this to arbitrary $\theta$ results in 
an effective sine-Gordon type theory of the form 
\begin{eqnarray}
{\cal S}_{meron}[\varphi]&=&\int d\tau dxdy [K(\partial_{\mu}\varphi)^2 +\lambda_{1}\cos\frac{\theta}{2}\cos\varphi
\nonumber \\
&&+\lambda_{2}\cos\theta\cos 2\varphi+
\cdot\cdot\cdot].
\end{eqnarray}    
In the above action, each harmonic term $(\propto\cos(n\varphi))$ represent processes 
associated with single, doubled, and tripled merons, etc. 
When $\theta\ne\pi$ ($\delta\ne 0$), the $\cos\varphi$ term is the 
most relevant (provided the fugacity expansion is valid) 
and picks out a unique ground state value for $\varphi$ with (depending on $\theta$) 
either $e^{i\varphi}=1$ or $e^{i\varphi}=-1$.   
Resorting to the symmetry analysis of Senthil and Fisher\cite{Senthil-Fisher}, one finds 
that this state corresponds to a vertical VBS state 
which simply follows the externally imposed bond-alternation pattern. 
The special role of the $\theta=\pi$ point 
manifests itself here in the vanishing of the $\cos\varphi$ term and all other odd harmonics. 
Here the doubled meron term 
$\cos 2\varphi$ becomes dominant, leading to doubly degenerate ground states satisfying 
$e^{i\varphi}=\pm i$\cite{Senthil-Fisher}. 
Hence, tracing the change of $\theta$ through $\theta=\pi$ shows that the system switches at this point 
between two vertical VBS states, corresponding to $e^{i\varphi}=1$ and $e^{i\varphi}=-1$ 
(Fig.2). 
\begin{figure}[htb]
\begin{center}
\end{center}
\caption{(Figure appended as a GIF file)The $\theta$-vacua for S=1/2 in the presence of AF-favoring anisotropy.}
\label{thetavacua}
\end{figure}%



We now discuss how the above $\theta$-vacua structure 
is likely to emerge in the context of actual spin systems.   
We know that 2d AFs have a much stronger tendency towards 
N{\'e}el ordering than in 1d, and one would generally expect an AF phase to intervene between 
different VBS phases\cite{Matsumoto-PRB-2001}. 
As briefly mentioned earlier, a partial clue regarding this question lies in the mapping process itself. 
Recalling that our low energy theory resulted from integrating out the subdominant modes 
$l_{\alpha}$, it is clear that the system consistent with the derivation should 
have such modes as the chief rapid fluctuations. Upon introducing a frustrating 
diagonal exchange term, configurations such as depicted in the lower half of 
Fig. 1 would start to have significant weight, thus weakening the AF dominance. 
(Similar situations may also arise when projecting 
a three dimensional anisotropic frustrated magnet onto an effective model of coupled chains
\cite{Starykh}.) 
If a direct inter-VBS transition is observed with an appropriate amount of frustration 
-perhaps supplemented with additional perturbations, 
our results suggest that 
such points are described by the strong coupling regime of our effective theory at $\theta=\pi$. 
It would also be interesting to relate our framework with large-N studies 
of frustrated magnets\cite{SR2}.   
Thus, although considerably fragile in nature compared to the 1d case, we believe the $\theta$-vacua 
of the 2+1d $O$(4) model should have relevance to actual frustrated magnets.        
     
We expect that   
various aspects of $\theta$-term physics in 1d will obviously find  
counterparts in our 2+1d problem. 
For instance, the $\theta$-term for an open spin chain contains a 
boundary contribution which induces a fractional-spin edge state\cite{Taikai}. 
This would correspond to decoupled spin-chain-like objects 
at the upper/lower ends of our system. 
One can also construct a 3d system with a $\theta$-vacua structure by 
repeating the above steps in one dimension higher.   
The appropriate starting point for this purpose would be the 
2+1d $O$(5) model which we derived in ref.[12], containing the 
WZ term 
\begin{eqnarray}
\Gamma^{2+1}[{\boldsymbol \phi}_{\rm VBS}, {\boldsymbol \phi}_{\rm AF}]&=&-i\frac{3}{4\pi}\int_{0}^{1}dt
\int d\tau dxdy\epsilon^{abcde}\nonumber\\
&&\phi_{a}\partial_{t}\phi_{b}\partial_{\tau}\phi_{c}
\partial_{x}\phi_{d}\partial_{y}\phi_{e},
\label{WZW-in-2+1}
\end{eqnarray}
 where the VBS portion of the five-component composite 
order parameter, ${\boldsymbol \phi}_{\rm VBS}$ now consists of 
two components. Comparing with eq.(\ref{wzw term1}), we see that  
this is the incarnation of the WZW term in 2+1d.  
Stacking up the 2d systems along a third ($z$-)direction, we obtain in the absence of bond-alternation 
a Berry phase term, $i\pi{\cal Q}_{\tau xyz}$, where 
${\cal Q}_{\tau xyz}
=
\frac{3}{8\pi^2}
\int 
d\tau dxdydz\epsilon^{abcde}
\phi_{a}\partial_{\tau}\phi_{b}\partial_{x}\phi_{c}\partial_{y}\phi_{d}\partial_{z}\phi_{e}.
$ 
This is again a NL$\sigma$ model with $\theta=\pi$, this time in 3+1d and on  
the target manifold $O$(5). 
It is easy to verify that adding on an interplanar bond-alternation would shift the value of $\theta$ 
in this case also, which may describe novel physics in 3d magnets.    
  
In summary we have shown that direct transitions among VBS states in anisotropic 2d AF systems 
can in certain cases be described in terms of the $\theta$-vacua structure of the 2+1d $O$(4) NL$\sigma$ model, in much 
the same way that inter-VBS transitions in spin chains have been understood using the $\theta$-term of 
the 1+1d $O$(3) NL$\sigma$ model. We suggested that realizations may be found in anisotropic frustrated 
magnets.  

We thank M. Kohno for fruitful discussions. 
This work was supported in part by Grant-in-Aid for Scientific Research (C) 10354143 from the 
Ministry of Education, Culture, Sports, Science and Technology of Japan.




\end{document}